\documentclass[iop,apj]{emulateapj} 

\usepackage{natbib}
\usepackage{graphicx}
\usepackage[usenames,dvipsnames]{color}

  \newcommand{\cmt} {~cm$^{-3}$} 

  \definecolor{gray}{RGB}{32,32,32}
  
\shortauthors{Motoyama et al.}

\begin{document}


\title{A Hydrochemical Hybrid Code for Astrophysical Problems. I. Code Verification and Benchmarks for Photon-Dominated Region (PDR)}


\author{Kazutaka Motoyama\altaffilmark{1,3}, Oscar Morata\altaffilmark{2},  Hsien Shang\altaffilmark{2,3,4},
Ruben Krasnopolsky\altaffilmark{2,3}, and Tatsuhiko Hasegawa\altaffilmark{2}}
\altaffiltext{1}{National Institute of Informatics, 2-1-2 Hitotsubashi, Chiyoda-ku, Tokyo 101-8430, Japan}
\altaffiltext{2}{Institute of Astronomy and Astrophysics, Academia Sinica, Taipei 10617, Taiwan}
\altaffiltext{3}{Theoretical Institute for Advanced Research in Astrophysics, Academia Sinica, Taipei 10617, Taiwan}
\altaffiltext{4}{Corresponding Author: shang@asiaa.sinica.edu.tw.}

%\altaffiltext{3}{Institute of Astronomy and Astrophysics, Academia Sinica,
%     Taipei 106, Taiwan.}


\begin{abstract}
A two dimensional hydrochemical hybrid code, {\it KM2}, is constructed to deal with astrophysical problems that would require coupled hydrodynamical and chemical evolution. 
The code assumes axisymmetry in cylindrical coordinate system, and consists of two modules: a hydrodynamics module and a chemistry module. The hydrodynamics module solves hydrodynamics using a Godunov-type finite volume scheme and treats included chemical species as passively advected scalars. The chemistry module implicitly solves non-equilibrium chemistry and change of the energy due to thermal processes with transfer of external ultraviolet radiation. Self-shielding effects on photodissociation of CO and H$_2$ are included.
In this introductory paper, the adopted numerical method is presented, along with code verifications using the hydrodynamics module, and a benchmark on the chemistry module with reactions specific to a photon-dominated region (PDR). Finally, as an example of the expected capability, the hydrochemical evolution of a PDR is presented based on the PDR benchmark.

\end{abstract}
    
\keywords{Astrochemistry --- Hydrodynamics --- Methods: numerical --- Photon-dominated region}

\section{Introduction}

%Physical Motivations
Spectral lines from molecules or ions in the interstellar medium can help to probe physical properties of the gas through density, temperature and kinematics. Chemical modeling of the interstellar medium is crucial to have accurate interpretations of astronomical observations as the intensities of spectral lines depend on spatial distribution and abundances of the individual species.  Recent advances in computing power also enable comprehensive radiative transfer simulations to predict a large number of observable spectral lines from theoretical and chemical models \citep[e.g.,][]{1997A&A...322..943J, 2000A&A...362..697H, 2010A&A...523A..25B}.  Next generation telescopes such as the Atacama Millimeter Array (ALMA) will significantly improve the sensitivity, angular and spectral resolutions of the observations. It is essential to establish accurate physical and chemical models of the interstellar media to derive physical properties of observed objects by comparing observations and theoretical models. 

%Star formation problems requiring hydrodynamics and chemistry
Chemical evolution of the interstellar medium is tightly coupled with its hydrodynamical evolution. The rates of chemical reactions depend sensitively on the density and temperature of the gas.
Molecular cloud cores with different magnetic field strengths, angular momenta, and initial density profiles have different hydrodynamical evolutions during gravitational collapse. Consequently, these cores are expected to have different chemical evolution as a result of the different distributions of density and temperature.
Recent observations reveal a variety of chemical properties in star forming regions that might have hinted such connections. For instance, depletion of CO is found at the dense interior of many starless cores, while L1521E does not show sign of depletion in spite of its central density similar to other objects \citep{2004A&A...414L..53T}. Some protostars harbor complex organic molecules \citep{2003ApJ...593L..51C}, others harbor carbon-chain molecules \citep{2008ApJ...672..371S}. 
These observed variations of chemical properties might be caused by their respective hydrodynamical evolution inside each of the cores.
Hydrochemical modeling of star formation process may help with the insights into the origins of the observed differences.

Many simulation works have been done to investigate chemical 
evolution in star forming cores \citep{2005ApJ...620..330A, 2009A&A...497..773V, 2012ApJ...758...86F}, 
circumstellar disks \citep{2009A&A...495..881V, 2011A&A...534A.132V}, and so on.
In these studies, chemical evolution was solved independently of hydrodynamical evolution using the 
density, temperature, and velocity obtained by hydrodynamic simulation or semi-analytic model.  Therefore, chemical evolution did not affect thermal and hydrodynamical
evolutions in their simulations. 
However, chemistry plays an important role in some astrophysical flows, because chemical 
abundance of the gas affects thermodynamics, and consequently dynamics,
of flows. For instance, atomic or molecular cooling is crucial for shock induced formation of 
molecular clouds \citep{2014ApJ...789...79A} and stars \citep{1998ApJ...508..291V}. 
As shown in this paper, photoevaporation flow of interstellar gas caused by strong external radiation 
field is also typical example in which chemical evolution affects thermal and hydrodynamical evolutions. 
In these situations, chemistry and hydrodynamics should be solved together in self-consistent way.

%Needs for such code
Most previous works with hydrodynamic simulations including chemical reactions have focused on evolution of primordial gas in the cosmological context. Chemistry is relatively simple in the primordial gas as it only involves about ten species consisting of primordial elements such as hydrogen, helium, and deuterium. On the contrary, chemical networks in the present-day interstellar medium are larger and more complex. To date, more than 170 interstellar molecules have been listed in the Cologne Database for Molecular Spectroscopy (CDMS\footnote{\url{http://www.astro.uni-koeln.de/cdms/molecules}}). 
Substantial efforts have been made to develop systematic solvers to follow the chemical evolutions of the species that would have played roles in the astrophysical problems of interests. For example, {\it Nahoon} (part of the "{\it KIDA}" package \citep{2012ApJS..199...21W}, Wakelam \& Herbst 2008), and {\it KROME} \citep{2014MNRAS.439.2386G} are two such packages that have incorporated chemistry solvers to solve large networks of chemical reactions. Hydrodynamic simulations aimed at the present-day universe would have to employ systematic solvers of this kind to properly tackle the problems. 

Hydrochemical simulations of present-day universe have to solve multi-physics characterized by various different time scales, namely hydrodynamics, chemical reactions, thermal processes, plus the radiative transfer. Hydrodynamics certainly affects chemical reactions in interstellar medium, while chemical reactions may play a role in the composition and thermal properties of the gas. Radiation may drive evolution of the interstellar medium through photoreactions and heating processes.
Although integrating all the physics directly with respect to time by their smallest time scale is straightforward and a simple way to achieve numerical stability and reliable results, long-term evolution is prohibited. It is necessary to establish an effective time integration method maintain both computational efficiency and accuracy. For these reasons, we developed a robust and computationally effective hydrochemical hybrid code, {\it KM2}, which we introduce in this paper.

%Why PDR Benchmarks. 
We will adopt models for the Photon-dominated regions (PDRs) as our benchmarks for capabilities developed with our hydrochemical hybrid {\it KM2} code for several reasons. 
First and foremost, the physical and chemical properties are dominated by penetrating far ultraviolet radiation (FUV: $6<h\nu<13.6$ eV), and their characteristics and emissions have been studied by a few code specifically developed for them: {\it Cloudy} \citep{1998PASP..110..761F, 2005ApJ...624..794S, 2005ApJS..161...65A}, {\it Meudon} \citep{1993A&A...267..233L, 2006ApJS..164..506L}, {\it UCL\_PDR} 
\citep{2002ApJ...579..270P, 2005MNRAS.357..961B}, {\it Leiden} \citep{1987ApJ...322..412B, 1988ApJ...334..771V, 1995A&A...302..223J}, {\it COSTAR} \citep{2000A&A...353..276K, 2001A&A...373..641K}, and {\it 3D-PDR} \citep{2012MNRAS.427.2100B}, and so on.
Comparisons and benchmark studies of these PDR codes were made by \citet{Rollig07} using the PDR models based on a common list of reaction networks. Although the hydrodynamics was not solved in these codes tested for the benchmarks,
it is useful to verify our solvers for chemical reactions, effect of radiative transfer, and thermal processes included. We first verified our code against these PDR benchmark models by performing chemical model calculations following similar criteria without actually advancing the hydrodynamics. Once the behaviors of the chemistry-only modules have been confirmed, the hydrodynamics module have also been tested following a few classical examples. A photoevaporating PDR model in 1D, with both of the hydrodynamics and chemistry modules active, was performed to showcase the potential dynamical effects from the hydrodynamics. To our knowledge, this is the first study of an evolving PDR model along with the impacts of its background thermal flow.

%Scopes covered in this paper. 
This paper aims to present an efficient approach to build a hydrochemical hybrid code based on existing hydrodynamic codes by applying modules of chemical solvers and reaction networks at their appropriate time steps.
Numerical method used in our hydrochemical hybrid code is described in detail.
The content is organized as follows. In Section \ref{sec: method}, 
we describe basic equations that we shall solve, design of the code, and details of the
numerical method. 
In Section \ref{sec: hd test}, hydrodynamics module of the code is verified against hydrodynamic tests.  
In Section \ref{sec: chem test}, chemical module of the code is verified against one dimensional PDR benchmarks in \citet{Rollig07} and multidimensional PDR model in \citet{2012MNRAS.427.2100B}.  
In Section \ref{sec: application}, we present a hybrid case with both of the hydro-and chemical modules turned on, for a model of PDR on top of background thermal flows. In Section \ref{sec: conclusions}, we discuss implications of the paper and future work.

%%%%%%%%%%%%%%%%%%%%%%%%%%%%%%%%%%%%%%%%%%%%%%%%%%%%%%%%%%%%%%%
%             numerical model
%%%%%%%%%%%%%%%%%%%%%%%%%%%%%%%%%%%%%%%%%%%%%%%%%%%%%%%%%%%%%%%

\section{Numerical Method}\label{sec: method}

\subsection{Basic Equations and Code Design}

  The hydrodynamic equations for the conservation of mass, momentum, and total energy are written as
  \begin{equation}
    \frac{\partial \rho}{\partial t} 
          +  \nabla \cdot \left( \rho \mbox{\boldmath $v$} \right) = 0,
  \label{eq: mass cons}
  \end{equation}
  
  \begin{equation}
    \rho \frac{\partial \mbox{\boldmath $v$} }{\partial t}
           + \rho \left( \mbox{\boldmath $v$} \cdot 
             \nabla \right) \mbox{\boldmath $v$}  = - \nabla P + \rho  \mbox{\boldmath $g$} ,
  \label{eq: momentum cons}
  \end{equation}
  
  \begin{equation}
    \frac{\partial E}{\partial t}
           + \nabla \cdot \left[ (E + P)  \mbox{\boldmath $v$} \right] 
            = \Gamma - \Lambda + \rho  \mbox{\boldmath $v$} \cdot \mbox{\boldmath $g$} ,
  \label{eq: energy cons}
  \end{equation}
where $\rho$, $\mbox{\boldmath $v$}$, $P$, $\mbox{\boldmath $g$}$ are the density, the velocity, the pressure, and the gravitational force, respectively. We include the heating $\Gamma$ and cooling $\Lambda$ rates in the energy equation. We adopt the equation of state for ideal gas, so that the total energy density of the gas is expressed as $E=P/(\gamma - 1) + \rho |\mbox{\boldmath $v$}|^2/2$, where $\gamma$ is the ratio of specific heats.

The chemical network includes the formation and destruction reactions of all the considered chemical species. The abundance of each species $i$ is controlled by a rate equation of the form
  \begin{eqnarray}
    \frac{{\rm d}n_i}{{\rm d}t} = \mathop{\sum}_{l}\zeta_{li}n_{l} +
     \mathop{\sum}_{j} \mathop{\sum}_{k} k_{jki} n_{j} n_{k} \nonumber \\ 
      - n_{i}\left (
     \mathop{\sum}_{l} \zeta_{il} + \mathop{\sum}_{l} \mathop{\sum}_{j}
     k_{\rm ijl} n_{j} \right ) 
    \label{eq:modelrate}
  \end{eqnarray}
  where $n_i$ is the number density of species $i$, $k_{jki}$ is the reaction rate coefficient for the reaction that form species $i$ from the reaction of species $j$ and $k$, $\zeta_{il}$ is the local photodestruction rate coefficient for the ionization or dissociation of species $i$ either by FUV photons or by cosmic ray induced photons, producing species $l$ as a result. 
The $\mu_{\mathrm{H}}$ denotes the mass per hydrogen nuclei whose typical value is $\sim$ 1.15 $m_{\mathrm{H}}$ in molecular cloud, where $m_{\mathrm{H}}$ is the mass of atomic hydrogen, and the number density of hydrogen nuclei $n_{\mathrm{H}}$ is related with the gas density $\rho$ by $n_{\mathrm{H}} = \rho / \mu_{\mathrm{H}} $.  

The operator-split method was adopted to solve hydrodynamical and chemical evolutions self-consistently. The basic equations are solved by the hydrodynamics and chemistry modules. In our approach, evolution of the energy due to cooling and heating processes is separately treated from evolution due to advection and external forces. That is, Equation (\ref{eq: energy cons}) is split into  
  \begin{equation}
    \frac{\partial E}{\partial t}
           + \nabla \cdot \left[ (E + P)  \mbox{\boldmath $v$} \right] 
            = \rho  \mbox{\boldmath $v$} \cdot \mbox{\boldmath $g$} ,
  \label{eq: energy cons mod}
  \end{equation}  
  and
  \begin{equation}
    \frac{\partial E}{\partial t} = \Gamma - \Lambda.
    \label{eq: thermal update}
  \end{equation}
The hydrodynamics module solves Equations (\ref{eq: mass cons}), (\ref{eq: momentum cons}), and (\ref{eq: energy cons mod}). The number density of each species $n_i$ is treated as passively advected variables in this step.
The chemistry module solves radiative transfer to obtain local mean intensity of FUV radiation, and then updates the number density of chemical species and the internal energy of the gas by solving Equations (\ref{eq:modelrate}) and (\ref{eq: thermal update}). The algorithm adopted to couple the hydrodynamics and the chemistry modules is described in detail in Section \ref{sec: time integration}.

%%%%%%%%%%%%%%%%%%%%%%%%%%%%%%%%%%%%%%%%%%%%%%%%%%%%%%%%%%%%%%%%%%%%%%%%%%%%%%%%%%
%      hydrodynamics module
%%%%%%%%%%%%%%%%%%%%%%%%%%%%%%%%%%%%%%%%%%%%%%%%%%%%%%%%%%%%%%%%%%%%%%%%%%%%%%%%%%

\subsection{The Module of Hydrodynamics}\label{sec: hd module}
  
The module of hydrodynamics solves Equations (\ref{eq: mass cons}), (\ref{eq: momentum cons}), and (\ref{eq: energy cons mod}).  The change of total energy of the gas due to thermal process is solved in chemistry module. In cylindrical coordinates system assuming axisymmetry, conservative form of equations can be written as  
  \begin{equation}
    \frac{\partial }{\partial t}  \mbox{\boldmath $U$}
                 + \frac{1}{r} \frac{\partial }{\partial r} r \mbox{\boldmath $F_r$} 
                 + \frac{\partial }{\partial z} \mbox{\boldmath $F_z$} =  
                 \mbox{\boldmath $S$} ,
    \label{eq: cylindrical}
  \end{equation}
  where   
  \begin{equation}
  \mbox{\boldmath $U$} = 
      \left(\rho,  
          \rho v_r, 
          \rho v_z, 
          \rho v_{\phi}, 
          E  
      \right)^T
  \end{equation}
  are conserved variables, 
  \begin{equation}
  \mbox{\boldmath $F_r$} =
      \left(
          \rho v_r, 
          \rho v_r^2 + P, 
          \rho v_z v_r, 
          \rho v_{\phi} v_r, 
          \left( E + P  \right) v_r 
      \right)^T
  \end{equation}
  and 
  \begin{equation}
  \mbox{\boldmath $F_z$} =
      \left(
          \rho v_z, 
          \rho v_r v_z,  
          \rho v_z^2 + P,  
          \rho v_{\phi} v_z, 
          \left( E + P  \right) v_z 
      \right)^T
  \end{equation}
  are its fluxes along r-direction and z-direction, respectively, 
  \begin{equation}
  \mbox{\boldmath $S$} =
      \left(
          0,  
          \rho g_r + \frac{\rho v_{\phi}^2 + p}{r},
          \rho g_z, 
           - \frac{\rho v_{\phi} v_r}{r}, 
          \rho  \mbox{\boldmath $v$} \cdot \mbox{\boldmath $g$}
      \right)^T
  \end{equation}
  are geometrical and physical source terms. These equations are integrated by Godunov-type finite volume scheme. 
An HLLC Riemann solver \citep{1994ShWav...4...25T} is used to compute flux at the cell interface. 
Data reconstruction scheme proposed by \citet{2014JCoPh.270..784M} is adopted to avoid large numerical errors near symmetry axis. Our hydrodynamics module has second-order accuracy in both space and time. Either uniform or non-uniform structured grid can be used. Non-uniform grid is advantageous when flow in specific region of 
computational domain needs to be solved with higher resolution than other regions.

The hydrodynamics module includes gravitational forces due to self-gravity of the gas and a specified point mass. Gravitational potential for the self-gravity $\Phi_{self}$ is obtained from Poisson's equation
  \begin{equation}
      \nabla^2 \Phi_{self} = 4 \pi G \rho,
  \label{eq: poisson}
  \end{equation}
where $G$ is the gravitational constant, and Equation (\ref{eq: poisson}) is solved by multigrid method \citep{1986nras.book.....P}. 
To reduce errors efficiently, iterative Poisson solver is run on the grid used for solving hydrodynamic equations and on coarser grids together. 
Boundary conditions are determined by following \cite{1996ApJ...468..784F}. The derivative $d \Phi/ d r$ is set to zero on z-axis. On the outer boundary, $\Phi_{self}$ is calculated
by Legendre polynomial expansion using the density distribution in the computational domain. For the point mass $\Phi_{pm}$, its potential is given by
  \begin{equation}
      \Phi_{pm} = - \frac{G M_{pm}}{\sqrt{r^2 + (z - z_{pm})}},
  \label{eq: point mass}
  \end{equation}
where $M_{pm}$ and $z_{pm}$ are the mass and z-coordinate of the point mass, respectively.
The point mass is located on the z-axis by axisymmetry. Using these gravitational potentials, the gravitational force is obtained as
  \begin{equation}
      \mbox{\boldmath $g$} = - \nabla (\Phi_{self} + \Phi_{pm} ).
  \end{equation}

%%%%%%%%%%%%%%%%%%%%%%%%%%%%%%%%%%%%%%%%%%%%%%%%%%%%%%%%%%%%%%%%%%%%%%%%%%%%%%%%%%
%      chemistry module
%%%%%%%%%%%%%%%%%%%%%%%%%%%%%%%%%%%%%%%%%%%%%%%%%%%%%%%%%%%%%%%%%%%%%%%%%%%%%%%%%%

\subsection{Chemistry Module}\label{sec: chem module}

%%%%%%%%%%%%%%%%%%%%%%%%%%%%%%%%%%%%%%%%%%%%%%%%%%%%%%%%%%%%%%%%%%%%%%%%%%%%%%%%%%
%      chemical network
%%%%%%%%%%%%%%%%%%%%%%%%%%%%%%%%%%%%%%%%%%%%%%%%%%%%%%%%%%%%%%%%%%%%%%%%%%%%%%%%%%

\subsubsection{Chemical Model and Thermal Evolution}

The module of chemistry solves a time-dependent set of equations for gas-phase chemistry based on the code used in \citet{Lee96} and \citet{MorataH08}. The original FORTRAN source code was modified into a subroutine that could be repetitively called from the main routine of {\it KM2}. The evolution of chemistry was tested to be the same for different physical conditions of density and temperature in the integrated
subroutine as in the original code.

The chemical solver involves a system of $N_s$ highly non-linear algebraic equations of the type in Equation (\ref{eq:modelrate}), one for each species.  The publicly available Fortran subroutine, DLSODE\footnote{\url{http://computation.llnl.gov/casc/software.php}}, which is based on the Gear method \citep{Gear71}, was adopted to find the time-dependent solutions. The Gear method is an implicit, linear multistep method that uses variable time steps and error control techniques to preserve the required accuracy
during the integration. Equations for chemical evolution are solved on the same grid as for 
hydrodynamic equations.

Arbitrary chemical reaction networks could be implemented through a Python routine in which different databases of chemical reactions could be read in. Recent databases for astrochemistry, such as UMIST\footnote{\url{http://www.udfa.net/}} \citep{2007A&A...466.1197W, 2013A&A...550A..36M}, and OSU\footnote{\url{http://www.physics.ohio-state.edu/\~{}eric/research.html}} \citep{2004MNRAS.350..323S, 2008ApJ...682..283G}, include several thousand
chemical reactions. The Python scripts read in  reaction data set written in the data format of UMIST or OSU, and generate Fortran subroutines of computing time derivatives for all included chemical species.

  The change of the energy due to the heating and cooling in reactions is implicitly computed by backward Euler method. Update of the energy from time $t$ to time $t + \Delta t$ is expressed as 
  \begin{equation}
     E(t + \Delta t) = E(t) + \Delta t \left[ \Gamma(t + \Delta t) - \Lambda(t + \Delta t) \right]. 
  \end{equation}
The cooling and heating rates are computed with chemical abundances at time $t + \Delta t$. The pressure and gas temperature are also updated using the updated energy. The gas temperature is given by 
  \begin{equation}
     T = \frac{\mu_{\mathrm{H}}[2 n(\mathrm{H_2}) + n(\mathrm{H}) + n(\mathrm{H^+})]}
              {n(\mathrm{H_2}) + n(\mathrm{H}) + n(\mathrm{H^+}) + n(\mathrm{e^-})} 
         \frac{P}{k_B \rho},
  \end{equation}
where $k_B$ is the Boltzmann's constant. Hereafter, $n(\mathrm{X})$ denotes the number density of species X. The mean molecular weight is obtained from number densities of $\mathrm{H_2}$, H, $\mathrm{H^+}$, and $\mathrm{e^-}$. 
 
%%%%%%%%%%%%%%%%%%%%%%%%%%%%%%%%%%%%%%%%%%%%%%%%%%%%%%%%%%%%%%%%%%%%%%%%%%%%%%%%%%
%      radiative transfer
%%%%%%%%%%%%%%%%%%%%%%%%%%%%%%%%%%%%%%%%%%%%%%%%%%%%%%%%%%%%%%%%%%%%%%%%%%%%%%%%%%

\subsubsection{Radiative Transfer and Photoreaction Rates}\label{sec: radiative transfer}
In {\it KM2}, transfer of the FUV radiation is solved to determine rates of photoreactions 
such as photoionizations and photodissociations.
The photoreaction rate due to FUV radiation is written as
  \begin{equation}
      k = \int_{\nu_t}^{\nu_{\mathrm{Lyl}}} \sigma_{\nu} \frac{4 \pi J_{\nu}}{h \nu} d \nu, 
  \end{equation}
where $\sigma_{\nu}$ is the cross section for reaction at the frequency $\nu$, $\nu_{\mathrm{Lyl}}$ is the Lyman limit frequency, $\nu_t$ is the threshold frequency for the reaction, $h$ is the Planck's constant, and $J_{\nu}$ is the local mean intensity of FUV radiation. Since a frequency-dependent radiative transfer is computationally too expensive to be coupled with hydrochemical evolution at each time step, we adopt a frequency-integrated FUV intensity normalized with the Draine standard radiation field \citep{1978ApJS...36..595D} 
to obtain photoreaction rates. \citet{1991ApJS...77..287R} showed that photoreaction rates of 
some species can be well fitted by function of visual extinction by using frequency dependent radiative transfer. We assumed that this approximation hold for photoreactions in our chemical network. 
  
The rate of the $i$th photoreaction in a semi-infinite plane-parallel geometry is given by the form of   
  \begin{equation}
k_i = \alpha_i \chi_0 \exp(- \gamma_i A_V)
  \end{equation}
where $\alpha_i$ is the photoreaction rate in the unshielded interstellar ultraviolet radiation field, $\chi_0$ is the scaling factor of the radiation field at cloud surface with respect to Draine standard radiation field, $A_V$ is the perpendicular visual extinction measured from cloud surface, and $\gamma_i$ is the parameter representing attenuation properties of the FUV. The visual extinction $A_V$ is related to hydrogen column density $N_{\mathrm{H}}$ by $A_V = 6.3 \times 10^{-22} N_{\mathrm{H}}$ \citep{1989MNRAS.237.1019W}.
  
For a cloud of arbitrary geometry, the photoreaction rates were computed in a similar manner as the three-dimensional PDR code developed by \cite{2012MNRAS.427.2100B}. Local intensity of the FUV radiation within a frequency range contributing the $i$th photoreaction at position ($r, z$) is calculated by averaging that of a semi-infinite plane-parallel cloud over all solid angle
\begin{equation}
      \chi_i(r,z) = \int_{0}^{4 \pi} \chi_0(\mathbf{q}) \exp(- \gamma_i A_V(\mathbf{q})) \frac{d \Omega}{4 \pi},
      \label{eq: chi average}
  \end{equation} 
where $\mathbf{q}$ is the vector oriented to the cloud surface with respect to the position ($r, z$). Photoreaction rate at the position ($r, z$) is given by
\begin{equation}
    k_i(r, z) = \alpha_i \chi_i(r, z).
\end{equation}
Since photodissociation of CO and $\mathrm{H_2}$ involves  line absorption, self- and mutual shielding effects are important, and shielding factors \citep{Lee96} are included in the integration of Equation (\ref{eq: chi average}) for these photoreactions.

%%%%%%%%%%%%%%%%%%%%%%%%%%%%%%%%%%%%%%%%%%%%%%%%%%%%%%%%%%%%%%%%%%%%%%%%%%%%%%%%%%
%      thermal processes
%%%%%%%%%%%%%%%%%%%%%%%%%%%%%%%%%%%%%%%%%%%%%%%%%%%%%%%%%%%%%%%%%%%%%%%%%%%%%%%%%%

\subsubsection{Heating Processes}

FUV radiation is main heating source in PDRs. 
FUV radiation ejects electrons from dust grains due to photoelectric effect. The kinetic energy of ejected electrons goes into the thermal energy of gas through interactions with ambient hydrogen molecules. The photoelectric heating rate is \citep{1994ApJ...427..822B}
  \begin{equation}
      \Gamma_{pe} = 10^{-24} \epsilon  G_0 n ~ \mathrm{ergs~ cm^{-3}~ s^{-1}}.
      \label{eq heat pe}
  \end{equation}
  Here, $G_0$ is the mean intensity of FUV radiation normalized with the Habing field \citep{1968BAN....19..421H}, which is related with the intensity normalized with the Draine field by $G_0 = 1.7 \chi_0$, and $n = n(\mathrm{H}) + 2n(\mathrm{H_2})$ is the number density of neutral hydrogen nuclei. The photoelectric heating efficiency $\epsilon$ is written as
  \begin{eqnarray}
    \epsilon &=& \frac{4.87 \times 10^{-2}}{1 + 4 \times 10^{-3} ( G_0 T^{1/2} /n(\mathrm{e^-}))^{0.73}} \nonumber \\
              && + \frac{3.65 \times 10^{-2} (T/10^4)^{0.7}}{1 + 2 \times 10^{-4} ( G_0 T^{1/2} / n(\mathrm{e^-}))}.
    \label{eq efficiency pe}
  \end{eqnarray}

  Line absorption of a FUV photon will pump $\mathrm{H_2}$ molecules to a bound excited electronic state. About 10 percent of the excited molecules drop back to the vibrational continuum of the ground electronic state and dissociates, and other 90 percent drop back to an excited vibrational state in the electronic ground state. Collisional de-excitation of FUV pumped $\mathrm{H_2}$ can be important heating source in dense gas.  
  Its heating rate is \citep{1979ApJS...41..555H}
  \begin{equation}
      \Gamma_{UV} =  9 R_d(\mathrm{H_2}) \frac{2.2}{1 + n_{cr}(\mathrm{H_2}) / n_{\mathrm{H}} } 
       ~ \mathrm{eV~ cm^{-3}~ s^{-1}},
  \end{equation}
  where $R_d(\mathrm{H_2})$ is photodissociation rate of molecular hydrogen. 
  The critical density is defined as
  \begin{eqnarray}
      n_{cr}(\mathrm{H_2}) & = & 10^6 T^{-1/2} / \{ 1.6 x_{\mathrm{H}} \exp [-(400/T)^2] \nonumber \\ 
             & & + 1.4 x_2 \exp{-[12000/(T + 1200)]} \} ~ \mathrm{cm^{-3}},
  \end{eqnarray}
  where $x_{\mathrm{H}}=n(\mathrm{H})/n_{\mathrm{H}}$ and $x_2 = n(\mathrm{H_2})/n_{\mathrm{H}}$ are fractional 
  abundance of atomic hydrogen and molecular hydrogen, 
  respectively.
  When the density exceeds the critical density, the excitation energy is effectively converted into 
  heat via collisional de-excitation rather than into radiation. 

  Photodestruction of atoms and/or molecules due to FUV radiation may also contribute to 
  the heating of gas. Photoionization of atomic carbon and photodissociation of molecular 
  hydrogen are considered in {\it KM2}. Kinetic energy of the ejected electron or H atom heats the gas. We adopted the value of 1.06 eV and 0.4 eV as released energies in a photoionization of atomic carbon \citep{2005A&A...436..397M} and in a photodissociation of molecular hydrogen \citep{1996A&A...307..271S}, respectively. 

  Cosmic rays become the main heating source where FUV radiation is strongly attenuated. Cosmic rays 
  ionize molecules and atoms in the cloud. The kinetic energy of the ejected electrons is converted to thermal energy of the gas. Ionization of molecular hydrogen
  \begin{equation}
      \mathrm{H_2} + \mathrm{cr} \rightarrow \mathrm{H} + \mathrm{H^+} + \mathrm{e^-} 
  \end{equation}
  \begin{equation}
      \mathrm{H_2} + \mathrm{cr} \rightarrow \mathrm{H_2^+} + \mathrm{e^-} 
  \end{equation}
  and ionization of atomic hydrogen
  \begin{equation}
      \mathrm{H} + \mathrm{cr} \rightarrow  \mathrm{H^+} + \mathrm{e^-} 
  \end{equation}
are considered as heating sources due to cosmic rays. With the cosmic ray ionization rates from \cite{1989ApJ...342..306H}, the cosmic ray heating rate is given by
  \begin{equation}
      \Gamma_{\mathrm{cr}} =  (0.952 n(\mathrm{H_2}) + 0.46 n(\mathrm{H})) \zeta_{\mathrm{cr}} 
       \Delta Q_{\mathrm{cr}}   ~ \mathrm{ergs~ cm^{-3}~ s^{-1}},
  \end{equation}
  where $\zeta_{\mathrm{cr}}$ is the total rate for electron production from cosmic ray ionization and 
  $\Delta Q_{\mathrm{cr}}$ is the energy deposited as heat as a result of the ionization. The values of $\zeta_{\mathrm{cr}} = 5.0 \times 10^{-17}$ $\mathrm{s^{-1}}$  and $\Delta Q_{\mathrm{cr}} = 20$ eV \citep{1978ApJ...222..881G} are adopted in the {\it KM2}.
  
  In interstellar clouds, molecular hydrogen is believed to form mainly on dust grains. 
  Formation rate of molecular hydrogen on dust grains is written as \citep{1985ApJ...291..722T}
  \begin{equation}
      R_f(\mathrm{H_2}) = 6 \times 10^{-17} (T/300)^{0.5} n(\mathrm{H}) n S(T),
  \end{equation}
  and it depends on the sticking coefficient of atomic hydrogen on dust grain
  \begin{equation}
      S(T) = \left[1 + 0.4(T + T_d)^{0.5} + 2 \times 10^{-3}T + 8 \times 10^{-6}T^2 \right]^{-1},
  \end{equation}
  where $T_d$ is the dust temperature determined by using the method in \cite{1991ApJ...377..192H}.
  This formation process yields 4.48 eV of binding energy, and 4.2 eV and 0.2 eV of this energy are assumed to have been distributed into vibrational/rotational excitation of the hydrogen molecule and thermal energy of the hydrogen molecule leaving the grain, respectively \citep{1979ApJS...41..555H}. 
Hence, heating rate due to formation of molecular hydrogen on grain surface is 
  \begin{equation}
%      \Gamma_{\mathrm{gr}} =  R_f \left\{ 0.2 + 4.2 [1 + n_{cr}(\mathrm{H_2}) / n_{\mathrm{H}} ]^{-1} \right\} 
      \Gamma_{\mathrm{gr}} =  R_f(\mathrm{H_2}) \left\{ 0.2 + \frac{4.2}{ 1 + n_{cr}(\mathrm{H_2}) / n_{\mathrm{H}} } \right\} 
      ~ \mathrm{eV~ cm^{-3}~ s^{-1}}.
  \end{equation}

\subsubsection{Cooling Processes}

  Collisional excitation of atomic fine-structure transition is an important source of radiative cooling in PDRs. The radiative cooling rates through the fine-structure lines of [\ion{O}{1}], [\ion{C}{1}], and [\ion{C}{2}] are calculated based on escape probability approximation \citep{1980A&A....91...68D,1985ApJ...291..722T}. The cooling rate due to a transition from level $m$ to level $n$ is given by
  \begin{equation}
      \Lambda_x(\nu_{mn}) = n_m A_{mn} h \nu_{mn} \beta_{esc}(\tau_{mn}) 
      \frac{S_x(\nu_{mn}) - P(\nu_{mn})}{S_x(\nu_{mn})}, 
  \end{equation}
  where $n_m$ is the population density in level $m$, $A_{mn}$ is the spontaneous transition probability, $\nu_{mn}$ is the transition frequency, $\beta_{esc}(\tau_{mn})$ is the escape probability of photon at optical depth $\tau_{mn}$, and $P(\nu_{mn})$ is the Planck function at the background temperature of 2.7 K. 
The source function is written as  
  \begin{equation}
      S(\nu_{mn}) = \frac{2 h \nu_{mn}^3}{c^2} \left( \frac{g_m n_n}{g_n n_m} - 1 \right)^{-1},
  \end{equation}
  where $c$ is the speed of light, $g_m$ and $g_n$ are the statistical weights of level $m$ and level $n$, respectively. Level population is determined by solving the equations of statistical equilibrium
  \begin{equation}
      n_m \sum^{l}_{n \neq m} R_{mn} =  \sum^{l}_{n \neq m} n_n R_{nm}, 
  \end{equation}
  where
  \begin{eqnarray}
  \begin{array}{ll} 
  R_{mn}   =  A_{mn} \beta_{esc}(\tau_{mn}) (1 + Q_{mn}) + C_{mn}, & (m > n) \\
  R_{mn}   =  (g_n / g_m) A_{nm} \beta_{esc}(\tau_{nm}) Q_{nm} + C_{mn}, &   (m < n)\\
  Q_{mn}=c^2 P(\nu_{mn}) / 2h\nu_{mn}^3
  \end{array}
  \end{eqnarray}
  Here, $C_{mn}$ is the collision-induced transition probability with atomic hydrogen and 
  molecular hydrogen. Atomic data for fine-structure transitions are taken from Leiden Atomic and Molecular Database (LAMDA\footnote{\url{http://home.strw.leidenuniv.nl/\~{}moldata/}},\citet{2005A&A...432..369S} and  summarized in Table \ref{table: atomic data}. Let $z_{\perp}$ denote the depth normal to plane, the optical depth and the escape probability for a plane-parallel cloud with uniform density can be expressed as \citep{1975ApJ...199...69D}
  \begin{equation}
      \tau_{mn}(z_{\perp}) = \frac{A_{mn} c^3}{8 \pi \nu_{mn}^3} \int_0^z n_m(z_{\perp}') 
                     \left[ \frac{n_n(z_{\perp}') g_m}{n_m(z_{\perp}') g_n} - 1 \right] \frac{dz_{\perp}'}{\delta \nu_d}
  \end{equation}
  and
  \begin{equation}
      \beta_{esc}(\tau) = \frac{1 - \exp(-3 \tau)}{3\tau},
      \label{eq: beta}
  \end{equation}
  respectively, where $\delta \nu_d$ is the root mean square of the thermal and turbulent velocities.
  We use the escape probability averaging that of Equation (\ref{eq: beta}) over all solid angles.

 \begin{table}
   \caption{Atomic data for fine-structure transitions}\label{table: atomic data}
   \centering
     \begin{tabular}{lccc}
     \tableline\tableline
           Species & Transition &  $\lambda$ ($\mathrm{\mu m}$)  &  A ($\mathrm{s^{-1}}$) \\
       \noalign{\smallskip}\hline\noalign{\smallskip}
       $\mathrm{C^+}$.....  & $^2P_{1/2}$ $-$ $^2P_{3/2}$ & 157.7 & $2.30 \times 10^{-6}$  \\
       C ......             &  $^3P_{0}$ $-$ $^3P_{1}$    & 609.1 & $7.88 \times 10^{-8}$  \\
                              &  $^3P_{1}$ $-$ $^3P_{2}$  & 370.4 & $2.65 \times 10^{-7}$ \\
                              &  $^3P_{0}$ $-$ $^3P_{2}$  & 230.3 & $1.81 \times 10^{-14}$ \\
       O ......             &  $^3P_{2}$ $-$ $^3P_{1}$    & 63.2  & $8.91 \times 10^{-5}$ \\
                              &  $^3P_{1}$ $-$ $^3P_{0}$  & 145.5 & $1.75 \times 10^{-5}$ \\
                              &  $^3P_{2}$ $-$ $^3P_{2}$  & 44.1  & $1.34 \times 10^{-10}$ \\
       \noalign{\smallskip}\hline\noalign{\smallskip}
     \end{tabular}
 \label{table_elemental}
 \end{table}
 
Free electrons and protons are important coolants in a highly ionized region. When a free electron recombines with a proton, the emitted photon removes internal energy of the gas. Cooling due to recombination is given as \citep{1963MNRAS.125..437H}
  \begin{equation}
      \Lambda_{rec} = \alpha_B  ( 1.09 + 0.158 \times 10^{-4} T  )
      k_B T n(\mathrm{e^-}) n(\mathrm{H^+})
  \end{equation}
  where $\alpha_B = 2.7 \times 10^{-13} \ \mathrm{cm^{-3}\, s^{-1}}$ is the case B recombination 
  coefficient for hydrogen. 
  When a free electron is accelerated by a proton, its kinetic energy can be removed by emitting a photon. 
  Cooling rate due to free-free emission is given by \citep{1989agna.book.....O}
  \begin{equation}
      \Lambda_{ff} = 1.42 \times 10^{-27} T^{1/2} g_{ff}  n(\mathrm{e^-}) n(\mathrm{H^+})
  \end{equation}
  where the Gaunt factor for free-free emission, $g_{ff}$, is assumed to be 1.3. 

  At the temperature of a few thousand Kelvin or higher, Ly$\alpha$ emission from atomic hydrogen and 6300 \AA \ line emission from atomic oxygen in the $^1D$ metastable level are effective cooling processes in atomic gas.
  The cooling rate due to Ly$\alpha$ radiation is given by \citep{1978ppim.book.....S}
  \begin{equation}
      \Lambda_{Ly\alpha} = 7.3 \times 10^{-19} n(\mathrm{e^-}) n(\mathrm{H}) e^{-118,400 / T}.
  \end{equation}
  The cooling rate due to [\ion{O}{1}]~6300 \AA \ emission is given by \citep{1985ApJ...291..722T}
  \begin{equation}
      \Lambda_{6300} = 1.8 \times 10^{-24} n(\mathrm{O}) [n(\mathrm{H}) + n(\mathrm{H_2})] 
      e^{-22,800/T}.
  \end{equation}

  Cool dust grains and carbon monoxide are important coolants in dense molecular region. The cooling rate of the gas by cooler dust grain is \citep{1989ApJ...342..306H}
  \begin{eqnarray}
      \Lambda_{d} &=& 1.2 \times 10^{-31} n^2 \left( \frac{T}{1000 \, \mathrm{K}} \right)^{1/2} 
                     \left( \frac{100 \, \mathrm{\AA}}{a_{\mathrm{min}}} \right)^{1/2} \nonumber \\
                   & \times &  \left[1 - 0.8 \exp(-75/T) \right] (T - T_d) ~ \mathrm{ergs~ cm^{-3}~ s^{-1}},
  \end{eqnarray}
  where $a_{\mathrm{min}}$ is the minimum size of grains. We adopt the value of 100 \AA{} for $a_{\mathrm{min}}$. Cooling due to CO rotational transitions is obtained from tabulated cooling functions for $T < 100$ K by \cite{1995ApJS..100..132N} and for $T>100$ K by \citet{1993ApJ...418..263N}. Optical depth averaged over all solid angles is used to obtain the cooling rate. 
  For cooling due to vibrational CO transitions, collisional excitation rate to $v=1$ state is taken from  
  \cite{1989ApJ...342..306H}. The cooling rates from collisions with H and $\mathrm{H_2}$ are
  \begin{eqnarray}
      \Lambda_{\mathrm{CO,vi}}^{\mathrm{H}} &=& 3.0 \times 10^{-12} \Delta E_{10}  T^{0.5}
                                           \exp \left[- \left( \frac{2000}{T} \right)^{3.43} \right] \nonumber \\
                                          & &  \times \exp \left( \frac{-3080}{T} \right) 
                                                n(\mathrm{CO}) n(\mathrm{H}) 
  \end{eqnarray}
  and
  \begin{eqnarray}
      \Lambda_{\mathrm{vib}}^{\mathrm{H_2}} &=& 4.3 \times 10^{-14} \Delta E_{10}  T
                                           \exp \left[- \left( \frac{3.14 \times 10^5}{T} \right)^{0.333}  \right] \nonumber \\
                                          & &  \times \exp \left( \frac{-3080}{T} \right) 
                                                n(\mathrm{CO}) n(\mathrm{H_2}), 
  \end{eqnarray}
  respectively, where $\Delta E_{10}=3080$ K $k_B$ is the energy of transition from $v=1$ to $v=0$ states.
  
%%%%%%%%%%%%%%%%%%%%%%%%%%%%%%%%%%%%%%%%%%%%%%%%%%%%%%%%%%%%%%%%%%%%%%%%%%%%%%%%%%
%      time integration
%%%%%%%%%%%%%%%%%%%%%%%%%%%%%%%%%%%%%%%%%%%%%%%%%%%%%%%%%%%%%%%%%%%%%%%%%%%%%%%%%%

\subsection{Time Integration}\label{sec: time integration}

In {\it KM2}, the modules of chemistry and hydrodynamics are coupled by operator splitting time integration method. In the hydrodynamics module, the time step of an update is determined by the Courant-Friedrichs-Lewy condition,
  \begin{equation}
    \Delta t_{hyd} =C_{CFL} \min \left( \frac{\Delta r}{c_s + |v_r|}, \frac{\Delta z}{c_s + |v_z|} \right),
  \label{dt hydro}
  \end{equation}
  where $C_{CFL}$ is a constant less than unity, $\Delta r$ and $\Delta z$ are widths of cells along the r and z directions, and $c_s$ is the sound speed. In the chemistry module, the number densities of chemical species and thermal energy are updated in a more complicated manner. Equation (\ref{eq:modelrate}) is solved under the assumption of constant gas temperature and shielding factors for photoreactions. Spatial distributions of $\mathrm{H_2}$ and CO affect their self-shielding. The timescale in which the self-shielding factors do not change significantly is given by 
  \begin{equation}
    \Delta t_{sh} =C_{sh} \min \left( \frac{n_\mathrm{H}}{| d \, n( \mathrm{H_2}) / dt|}, 
    \frac{X_{\mathrm{C}} n_{\mathrm{H}}}{|d \, n( \mathrm{CO})  / dt|} \right),
  \label{dt sh}
  \end{equation}
  where $C_{sh}$ is a constant of order unity and $X_{\mathrm{C}}$ is the elemental abundance of carbon. The value of $C_{sh}$ is typically set to be from 0.1 to 1.0. Number densities of the chemical species are updated with the time step $\Delta t_{sh}$, then thermal energy and temperature are updated using the cooling and heating rates computed with the updated chemical abundances. If the relative change of temperature from the previous step is larger than the tolerance value $C_{th}$, the chemical and thermal updates will be redone with a smaller time step.  The value of $C_{th}$ is set to be less than 0.1 in the benchmarks. The thermal and chemical updates are sub-cycled relative to the single hydrodynamic update, since timescales for changes of temperature and shielding factors are smaller than $\Delta t_{hyd}$ in general. 
  
  \begin{figure}
  \begin{center} 

  \includegraphics[width=\columnwidth]{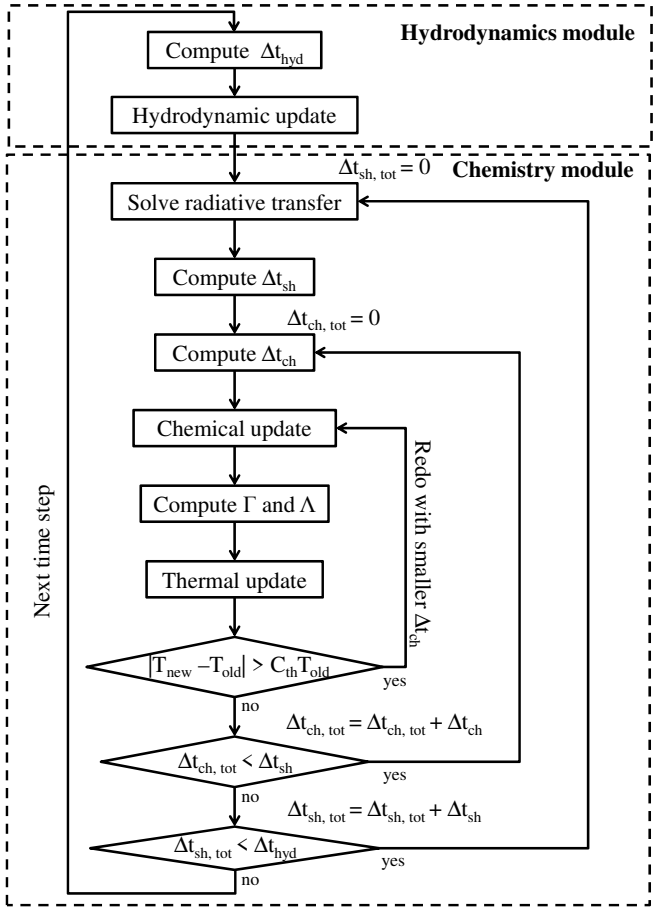}

  \end{center}
  \caption{Schematic diagram of the algorithm adopted. The hydrodynamics module and the chemistry module are boxed by the dashed lines, and the control loops are connected by solid arrows.}
  \label{fig flowchart}
 \end{figure} 
  
  Figure \ref{fig flowchart} shows the schematics of our time integration algorithm in block form. 
  The procedure to update physical quantities from time $t$ to $t + \Delta t_{hyd}$ is summarized as follows. 
  \begin{enumerate}
      \item First, the time step $\Delta t_{hyd}$ for hydrodynamic update is computed from Equation (\ref{dt hydro}). 
      \item The physical quantities $\rho$, \mbox{\boldmath $v$}, $P$, and $E$ are updated with the time step $\Delta t_{hyd}$ by solving Equation (\ref{eq: cylindrical}).
      \item\label{step RT} The radiative transfer is solved to obtain FUV intensity and shielding factors for photoreactions. 
      \item The time step for updating shielding factors, $\Delta t_{sh}$ is computed from Equation (\ref{dt sh}).
      \item\label{step dt_ch} The time step for chemical and thermal update, $\Delta t_{ch}$, is computed from 
            $\Delta t_{ch} = \Delta t_{sh} - \Delta t_{ch, tot}$, 
            where $\Delta t_{ch,tot}$ is the cumulative total of $\Delta t_{ch}$. 
      \item\label{step chemical update} The number densities of included chemical species $n_i$ are updated by solving Equation (\ref{eq:modelrate}). The temperature and shielding factors are kept constant.
      
      \item The heating $\Gamma$ and cooling $\Lambda$ rates are computed with the updated chemical abundances.
      \item\label{step thermal} Then $E$ and $T$ are updated implicitly. If the relative change of the temperature from previous step is larger than $C_{th}$, go back to step \ref{step chemical update} and redo the following steps with smaller $\Delta t_{ch}$. 
      \item\label{step last ch} Steps \ref{step dt_ch} through \ref{step thermal} are repeated until 
            $\Delta t_{ch,tot}$ becomes equal to $\Delta t_{sh}$.
      \item Steps \ref{step RT} through \ref{step last ch} are then repeated until $\Delta t_{sh, tot}$ becomes equal to $\Delta t_{hyd}$, where $\Delta t_{sh, tot}$ is the cumulative total of $\Delta t_{sh}$.
  \end{enumerate}
  
  Equations (\ref{dt hydro}) and (\ref{dt sh}) are computed over all the computational cells, and their minimum values are adopted as time steps $\Delta t_{hyd}$ and $\Delta t_{sh}$. 
  By contrast, the time step $\Delta t_{ch}$ is locally determined at each cell. 
  It allows us to use large time step in cells where the gas is close to thermal equilibrium, and reduce computational time compared to using a global time step.

%%%%%%%%%%%%%%%%%%%%%%%%%%%%%%%%%%%%%%%%%%%%%%%%%%%%%%%%%%%%%%%%%%%%%%%%%%%%%%%%%%
%      code tests
%%%%%%%%%%%%%%%%%%%%%%%%%%%%%%%%%%%%%%%%%%%%%%%%%%%%%%%%%%%%%%%%%%%%%%%%%%%%%%%%%%

\section{Tests for the Hydrodynamics}\label{sec: hd test}

The hydrodynamics module has been tested with some standard hydrodynamical problems with known analytic solutions such as the Sod shock tube and the Sedov solution.  The chemistry module was turned off during these tests. The Sod shock tube \citep{1978JCoPh..27....1S} is a Riemann problem to test accuracy and ability of computational hydrodynamics code in simulating compressible flow with shock wave. 
High pressure gas and low pressure gas were initially separated by contact discontinuity and at rest everywhere.
The initial density and pressure were unity at $z < 0.5$, while 0.125 and 0.1 at $z > 0.5$. 
The ratio of specific heats $\gamma$ was chosen to be 1.4.
The size of the computational domain in z was unity and covered with 150 cells.  

\begin{figure}
  \begin{center} 

  \includegraphics[width=\columnwidth]{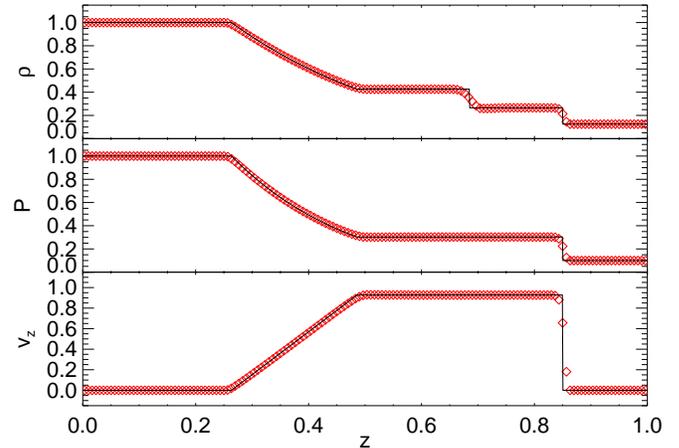}

  \end{center}
  \caption{The density (top), the pressure (middle), and the velocity (bottom) profiles for Sod shock tube test 
  at $t=0.2$. 
  The diamonds represent numerical results. Analytic solutions are plotted by solid lines as references.}
  \label{fig Sod}
 \end{figure}

Figure \ref{fig Sod} shows profiles of the density, the pressure, and the velocity along z-axis at the time $t = 0.2$. 
Sod shock tube involves three type nonlinear waves, that is, shock wave, contact discontinuity, and rarefaction wave.
The shock wave and the contact discontinuity propagate to the right, and the rarefaction wave propagates to the left. 
Positions of these three waves are identical to those of the analytic solution within resolution. 
The shock front is located at $z=0.85$, and it is resolved with a few cells without post-shock numerical oscillation.
The contact discontinuity can be seen at $z=0.68$ only in the density profile. 
The rarefaction front is located at $z=0.26$. The density, the pressure, and the velocity behind it show good 
agreements with analytic solutions.  

As a multidimensional test, we show results of point-like explosion in homogeneous medium, which is known as 
the Sedov explosion problem.
Sedov explosion involves self-similar propagation of a strong spherical shock wave through the 
background homogeneous medium. 
The exact analytic solution is given by \cite{1959sdmm.book.....S}.
Our computational domain extended from 0 to 20 pc in z with 1024 cells, and from 0 to 10 pc in r with 512 cells. 
The initial number density $n_{\mathrm{H}}$ was set to be 1 $\mathrm{cm^{-3}}$. 
We deposited a quantity of thermal energy $E=10^{51}$ erg into a central small region whose radius is 0.4 pc 
to initiate the explosion. 
The ratio of specific heats $\gamma$ was chosen to be 1.4.

  \begin{figure}
  \begin{center} 

  \includegraphics[width=\columnwidth]{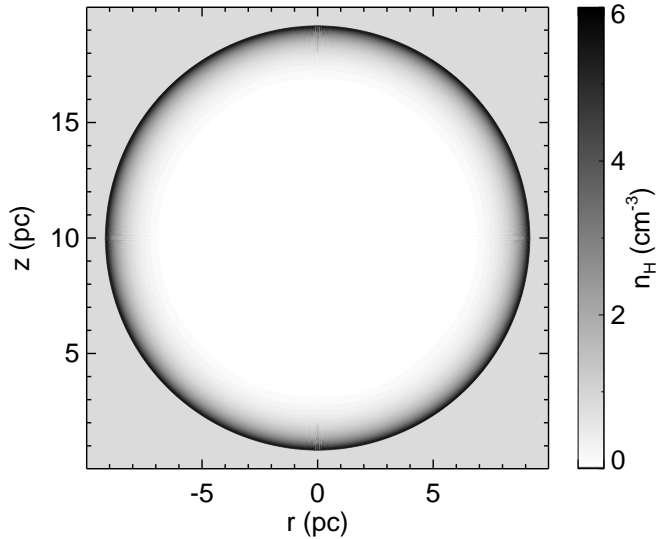}

  \end{center}
  \caption{The number density distribution  
  at $t=6$ kyr for Sedov explosion test. }
  \label{fig Sedov density}
 \end{figure}

 \begin{figure}
  \begin{center} 

  % color figure for online version
  \includegraphics[width=\columnwidth]{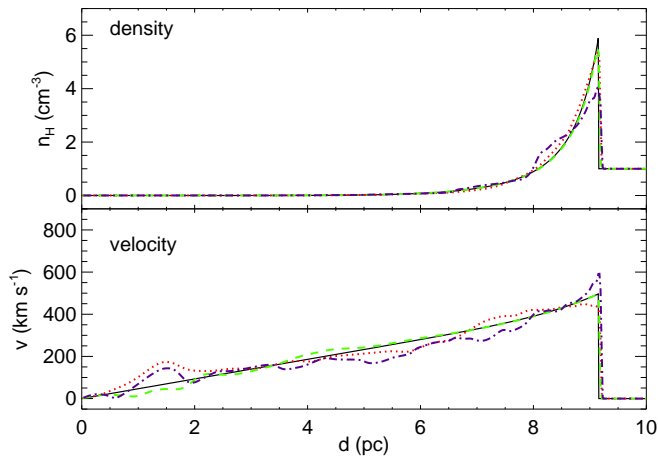}

  \end{center}
  \caption{The number density and the velocity profiles for Sedov explosion test at $t=6$ kyr
  as functions of the distance from the center of explosion. The dashed (green), the dotted (red), and 
  the dash-dotted (purple) lines represent profiles along directions at angle 0, 45, and 90  degree with r-axis, respectively. 
  The solid lines represent analytic solutions. }
  \label{fig Sedov profiles}
 \end{figure}

Figure \ref{fig Sedov density} shows the number density field at $t=6$ kyr. 
For presentation, the figure is symmetrized with respect to the z-axis.
Shock wave spherically propagated, and good symmetry was maintained in the density field.
We confirmed that the total energy and total mass of the system were conserved within errors of $2.5 \times 10^{-11}$ percent and 
$1.1 \times 10^{-11}$ percent through the simulation, respectively. 
Figure \ref{fig Sedov profiles} shows density and velocity profiles along different directions from the center 
of explosion with exact analytic solutions. 
Shock front is sharply resolved, and its position is identical to analytic solution.  
Although the velocity profile along z-direction slightly deviates from the analytic 
solution, numerical results show good agreements with analytic solutions on the whole.
The discrepancy between numerical and analytic solutions near z-axis is attributed to errors from geometrical source term $\frac{p}{r}$ in Equation (\ref{eq: cylindrical}).
Sedov explosion is the case that errors from geometrical source term $\frac{p}{r}$ severely affect
results, because high pressure gas was set at small central region as initial condition. 
In most other cases, errors from geometrical source term are negligible.

\section{Tests for the Chemistry module}\label{sec: chem test}

Similar benchmark tests have been conducted for the chemistry module with the hydrodynamics module turned off.
In this context, the PDR benchmarks presented in \citet{Rollig07} have been chosen to be the references.
These tests mainly show how well the included chemical network based on the PDR models and the specific cooling and heating processes are doing in comparison with other
codes and their respective chemistry and physical processes designed to model PDRs. These tests (in the context of \citet{Rollig07}) use a reduced chemical network only with some of the more fundamental molecules and reactions that had been previously adopted in the joint 
benchmark effort \citet{Rollig07} by several different groups. 
In addition to these one dimensional PDR benchmark tests, application example of ``{\it 3D-PDR}'' 
presented by \citet{2012MNRAS.427.2100B} has been chosen for test of ability to solve multidimensional problems.   
Chemical and thermal structures of spherical cloud illuminated by FUV radiation has been solved 
in this test.

\subsection{The PDR benchmarks}\label{sect:pdr_benchmark}

The reaction network adopted exactly identical to the one in \citet{Rollig07} prepared for the benchmarking of PDR models includes the four most abundant elements (H, He, O, and C) for a total of 31 species. The reaction rates were taken from the UMIST99 database, with some corrections, for a total of 287 reactions. Table~\ref{table_elemental} gives the initial abundances of the species used, in which elemental cosmic abundances for H, He, C, and O were assumed.

 \begin{table}
   \caption{Atomic initial abundances}
   \centering
     \begin{tabular}{ll}
       Element & n(X)/n$_{\rm H}$ \\
       \noalign{\smallskip}\hline\noalign{\smallskip}
       H     & 1\\
       He    & \hspace{0mm}0.1 \\
       C     & $1\times10^{-4}$\\
       O     & $3\times10^{-4}$\\
       \noalign{\smallskip}\hline\noalign{\smallskip}
     \end{tabular}
 \label{table_elemental}
 \end{table}

The eight test cases discussed in \citet{Rollig07} were conducted, which covered number densities $n_{\mathrm{H}}$ of $10^3$ and $10^{5.5}$\cmt, and the impinging radiation fields with FUV intensities $\chi$ of 10 and $10^5$ in units of the Draine field. There were also two sets of models: the ``F models'', in which a fixed gas temperature of 50~K and dust temperature of 20~K were adopted, and the ``V models'', in which the gas and dust temperature along the PDR were self-consistently calculated.

\subsubsection{The F Models}

\begin{figure*}
   \begin{center}
   % color figure for online version
   \includegraphics[width=0.7 \textwidth]{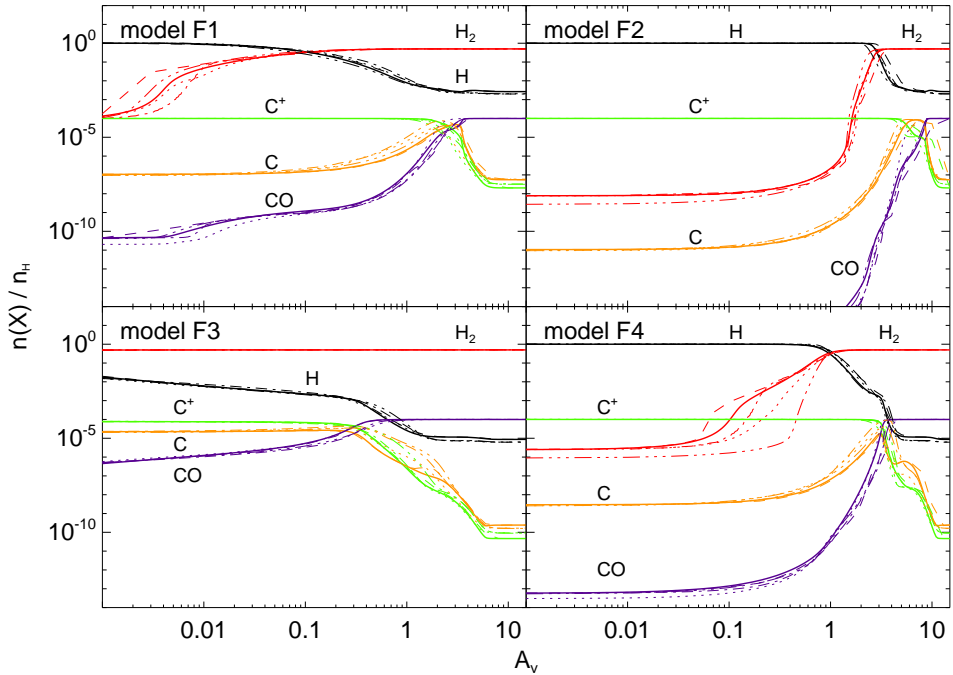}
   
   % black and white figure for print version
   %\includegraphics[width=0.7 \textwidth]{f5.eps}
   \end{center}
   \caption{Chemical abundances as a function of visual extinction for the benchmark models with a fixed temperature of 50 K from \citet{Rollig07} and the {\it KM2} code. The thick solid lines show the abundances of H (black), H$_2$ (red), C$^+$ (green), C (orange), and CO (purple) from the {\it KM2}, and results from other models are labeled as ``{\it Lee96mod}'' \textit{(dashed lines)}, ``{\it Leiden}'' \textit{(dotted lines)}, ``{\it UCL\_PDR}'' \textit{(dot-dashed lines)}, ``{\it Meudon}'' \textit{(three-dots-dashed lines)}. \textit{Upper row)} Model F1: density $n_{\mathrm{H}}=10^3$\cmt, FUV radiation field intensity $\chi=10$ in units of the Draine field; Model F2: $n_{\mathrm{H}}=10^3$\cmt, $\chi=10^5$. \textit{Bottom row)} Model F3: density $n_{\mathrm{H}}=10^{5.5}$\cmt, FUV radiation field intensity $\chi=10$; Model F4: $n_{\mathrm{H}}=10^{5.5}$\cmt, $\chi=10^5$.}
 \label{Fmodels}
 \end{figure*}

Figure~\ref{Fmodels} shows the comparison of results from the {\it KM2} code with ``{\it Lee96mod}'', a time-dependent code which uses the same treatment as \citet{MorataH08} and this paper, ``{\it UCL\_PDR}'', another time-dependent code, ``{\it Leiden}'' and ``{\it Meudon}'', for the case of a fixed gas temperature, $T=50$~K, the "F" Models, from \citet{Rollig07}.

The results are practically identical for most of the visual extinction values in the four "F" Models. When the cases of {\it KM2} differ somewhat more from others, there is a wider spread of results among the models. At very low extinctions for the abundance of H$_2$ for the F1 model or at $A_V\sim0.1$ for the F4 model, the model predictions scatter while agree with one another reasonable well otherwise. For these cases, {\it KM2} tends to fall within the spread of abundances. The differences could be attributed to differences in the ways of calculating the H$_2$ and CO photodissociation rates among the codes and not to the way that {\it KM2} propagates the radiation from the outer to the inner layers.

\subsubsection{The V Models}\label{sec V model}

\begin{figure*}
   \begin{center}
   % color figure for online version
   \includegraphics[width=0.7 \textwidth]{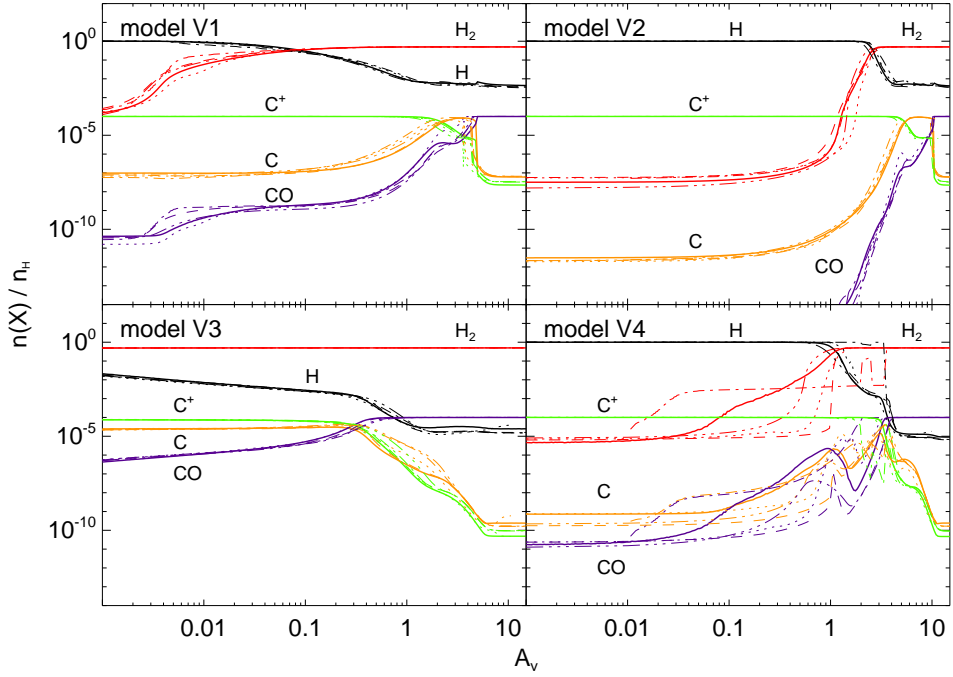}

   % black and white figure for print version
   %\includegraphics[width=0.7 \textwidth]{f6.eps}
   \end{center}
   \caption{The "V" models of the chemical abundances as a function of visual extinction with a variable temperature solved through the heating and cooling processes for the benchmark models from \citet{Rollig07} and the {\it KM2} code. The thick solid lines show the abundances of H (black), H$_2$ (red), C$^+$ (green), C (orange), and CO (purple) from the {\it KM2}, and results from other models are labeled as ``{\it Lee96mod}'' \textit{(dashed lines)}, ``{\it Leiden}'' \textit{(dotted lines)}, ``{\it UCL\_PDR}'' \textit{(dot-dashed lines)}, ``{\it Meudon}'' \textit{(three-dots-dashed lines)}. \textit{Upper row)} Model V1: density $n_{\mathrm{H}}=10^3$\cmt, FUV radiation field intensity $\chi=10$ in units of the Draine field; Model V2: $n_{\mathrm{H}}=10^3$\cmt, $\chi=10^5$. \textit{Bottom row)} Model V3: density $n_{\mathrm{H}}=10^{5.5}$\cmt, FUV radiation field intensity $\chi=10$; Model V4: $n_{\mathrm{H}}=10^{5.5}$\cmt, $\chi=10^5$.}
 \label{Vmodels}
 \end{figure*}

\begin{figure}
   % color figure for online version
   \includegraphics[width=\columnwidth]{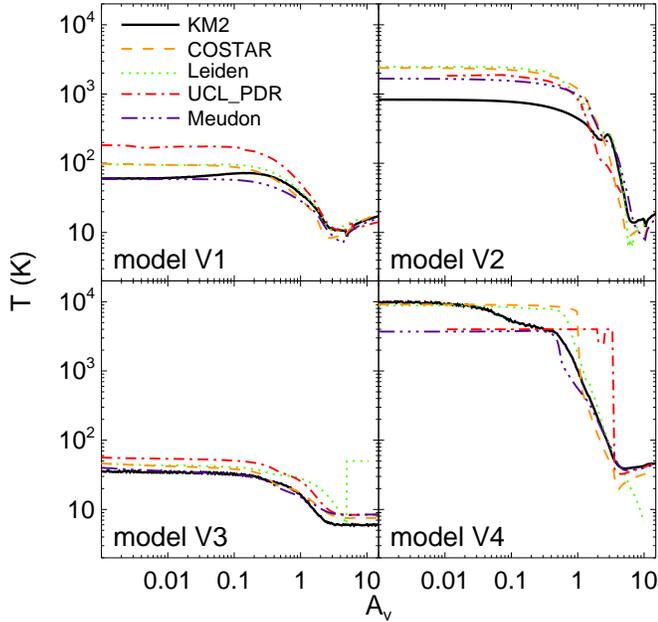}

   % black and white figure for print version
   %\includegraphics[width=\columnwidth]{f7.eps}
   \caption{Temperature as a function of visual extinction for the benchmark models and the {\it KM2} code (thick black line). Other codes are labeled as ``{\it COSTAR}'' \textit{(dashed lines)}, ``{\it Leiden}'' \textit{(dotted lines)}, ``{\it UCL\_PDR}''\textit{(dot-dashed lines)}, ``{\it Meudon}'' \textit{(three-dots-dashed lines)}.
   }
 \label{figure_T}
 \end{figure}

Figure~\ref{Vmodels} shows the comparison of the {\it KM2} code for the case where the gas temperature is self-consistently calculated from the cooling and heating functions included in the codes. The cases taken from \citet{Rollig07} are ``{\it UCL\_PDR}'', ``{\it Leiden}'', ``{\it COSTAR}'', and ``{\it Meudon}''. The gas temperature will be different at each layer of the PDR.
Figure~\ref{figure_T} shows the distribution of the gas temperature as a function of the visual extinction into the cloud for the codes shown in Figure~\ref{Vmodels}.

Figure~\ref{Vmodels} illustrates the models with variable temperature tend to show a significantly larger spread in the results among the codes. Still the {\it KM2} in general agrees very well with the rest of the codes, for the ranges of $A_V$, and for the models V1 to V3. They appear very similar to those found in the cases in the "F" models. On the other hand, the models vary greatly for some of the molecules (H$_2$, CO) in the V4 models, which have lower agreements among the participating codes in the range of $A_V=0.01$ to 3. {\it KM2} is not much worse than others in this regard.

In the models V1 to V3, the gas temperatures calculated by the {\it KM2} code are lower by a factor of a few for $A_V\la1$. This may be due to the larger cooling rate of atomic oxygen in {\it KM2}, whose collisional excitation rates of atomic oxygen were taken from \citet{2007ApJ...654.1171A}. These excitation rates are larger than values used in other codes previously by a factor of a few, which were taken from \citet{1977A&A....56..289L}. The cooling in \ion{O}{1} is more effective than those of other codes. The V4 model shows again the largest difference in temperatures, and the temperature predicted by the {\it KM2} code is within the spread of temperatures from other codes. These all indicate that performance based on the PDR specific problem and the specific set of reaction rates adopted in \citet{Rollig07} work well the overall design of the {\it KM2}. In next section, we discuss a sample model of its future use.

\subsection{The spherical cloud illuminated by a plane-parallel FUV radiation}
In this section, we present chemical and thermal structures of a spherical cloud illuminated 
by a plane-parallel FUV radiation.  
The spherical cloud had uniform density of $n_{\mathrm{H}}=10^3$ $\mathrm{cm^{-3}}$ and 
a radius of 5.15 pc. 
Our computational domain extended from 0 to 10.3 pc in z with 2048 cells, and from 0 to 5.15 pc 
in r with 1024 cells. 
A center of the cloud was located at $r=0$ pc and $z=5.15$ pc.
The incident FUV radiation with intensity of $\chi=10$ entered from bottom boundary of the computational domain.
The density and intensity of the incident FUV radiation were same as model V1 shown 
in Section \ref{sec V model}.
Chemical network and initial chemical abundances used in this test were also same as PDR benchmark tests. 
The evolution was solved for 30 Myr. It was long enough to achieve chemical and thermal 
steady state.

\begin{figure}[htb]
   \begin{center}
   \includegraphics[width=\columnwidth]{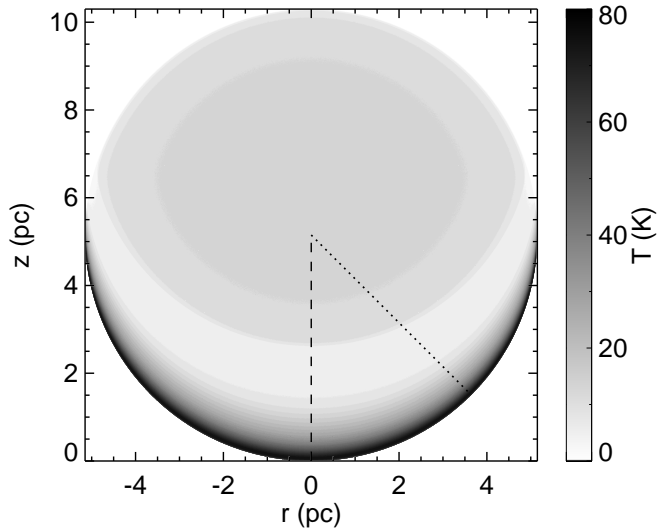}
   \end{center}
   \caption{The temperature distribution of spherical cloud illuminated by a plane-parallel FUV radiation. The dashed and dotted lines 
   represent directions along which chemical abundances plotted in Fig. \ref{figure abundance_sphere}.}
 \label{figure tempe_sphere}
 \end{figure}

\begin{figure}[htb]
   \begin{center}
   \includegraphics[width=0.8 \columnwidth]{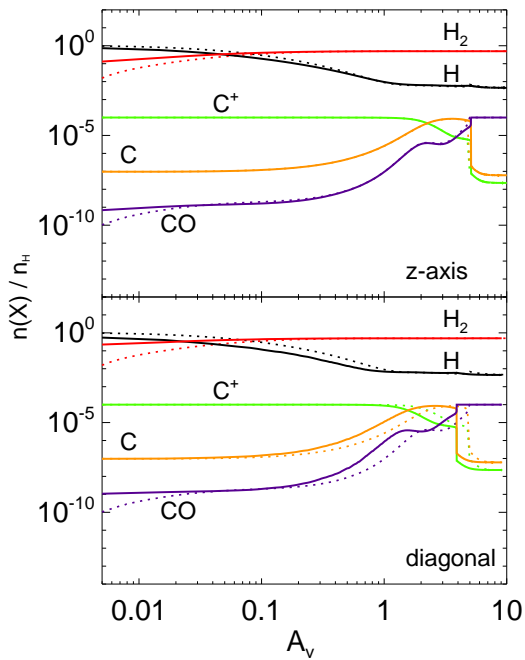}
   \end{center}
   \caption{The chemical abundances as a function of visual extinction along 
   z-axis (dashed line in Fig. \ref{figure tempe_sphere}) and diagonal direction (dotted line in Fig. \ref{figure tempe_sphere}). The dotted lines represent results of model V1 in Section \ref{sec V model}.}
 \label{figure abundance_sphere}
 \end{figure}

Fig. \ref{figure tempe_sphere} shows the temperature distribution of the cloud. 
For presentation, the figure is symmetrized with respect to the z-axis.
The cloud surface of the bottom semi-sphere is heated to $\simeq 80$ K by heating due to 
FUV radiation. As the FUV radiation is attenuated, its heating becomes inefficient.  
The temperature decreases to $\simeq 10$ K within thin region of $\la 1$ pc near the surface. 
The inner region of the upper semi-sphere has slightly higher temperature than outer region, 
because the efficiency of the cooling is lower there due to larger optical depth. 
The temperature profile along z-axis shows good agreement with results of model V1, 
the deviation between them is within a few K. 
Fig. \ref{figure abundance_sphere} shows the chemical abundances as a function of visual 
extinction along z-axis and a diagonal direction with 45 degrees respect to z-axis. 
The chemical abundances along z-axis show good agreements with those of model V1.
On the other hand, intensity of FUV radiation decreases more steeply along the diagonal direction. 
All transitions of H/$\mathrm{H_2}$, $\mathrm{C^+}$/C, and C/CO occur at smaller visual 
extinction than model V1.

Our results are consistent with those of ``{\it 3D-PDR}'' presented in \citet{2012MNRAS.427.2100B}, 
except the temperature of surface region at the bottom semi-sphere is lower than $\simeq 170$ K 
of ``{\it 3D-PDR}''. 
``{\it 3D-PDR}'' is an extension of ``{\it UCL\_PDR}'' for three dimensional models, therefore 
both codes produce identical results in one dimensional PDR benchmarks. 
As seen in Fig. \ref{figure_T}, these codes tend to produce higher temperature than other PDR 
codes for models with FUV radiation intensity of $\chi=10$.

%%%%%%%%%%%%%%%%%%%%%%%%%%%%%%%%%%%%%%%%%%%%%%%%%%%%%%%%%%%%%%%%%%%%%%%%%%%%%%%%%%
%      applications
%%%%%%%%%%%%%%%%%%%%%%%%%%%%%%%%%%%%%%%%%%%%%%%%%%%%%%%%%%%%%%%%%%%%%%%%%%%%%%%%%%

\section{Applications}\label{sec: application}

As a test example for the more general astrophysical problems, the hybrid hydrochemical evolution of a PDR is presented here for the intended capabilities of the {\it KM2} code. The initial density and intensity of FUV radiation field were set to be the same as the PDR benchmark model V2, namely $n=10^3$ and $\chi = 10^5$. The chemical network and initial chemical abundance were also the same as the benchmark models in Section \ref{sec: chem test}. Our computational domain was one-dimensional and extended from 0 to 7 pc in z with 1024 uniform cells. 
As a reference, simulation with larger chemical network has also been performed.  
For the reference model, 2799 reactions composed of H, He, O, and C were taken from UMIST RATE12 database.
The reaction rates for formation and photodissociation of H$_2$ were same as PDR benchmark tests.
The chemical network for the reference model include 2801 reactions among 203 species in total.
This reference model was intended to see how difference of implemented chemical network affects hydrochemical evolution. 
  
\begin{figure}[htb]
   \begin{center}
   \includegraphics[width=\columnwidth]{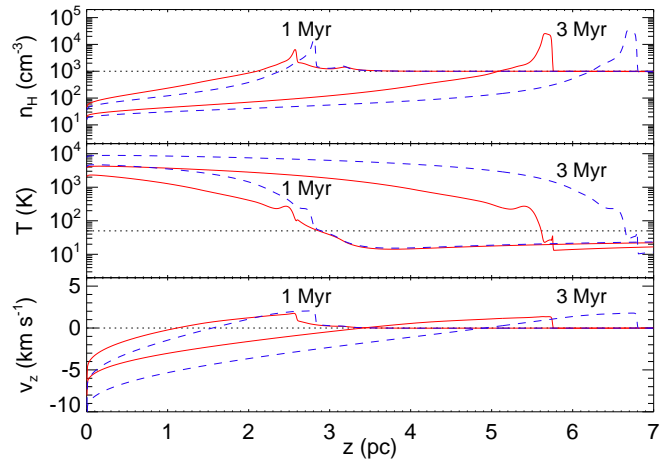}
   \end{center}
   \caption{Number density, temperature, and velocity profiles as a function of z-coordinates 
   in our hydrochemical simulations. 
   The solid and the dashed lines represent results for PDR benchmark model V2 and for the reference model with larger chemical network, respectively. The dotted lines represent profiles at $t=0$.}
 \label{figure pdr_hydro}
 \end{figure}

Figure \ref{figure pdr_hydro} shows the time evolution of the number density, temperature, and velocity profiles. The incident FUV radiation enters from the left boundary and heats up the gas to $\sim$ 4000 K at the surface. The expansion of the hot gas generates an evaporation flow that accelerates leftward at a velocity of $\sim$ 6 km s$^{-1}$. 
At the same time, a back-reaction of the photoevaporation flow drives a shock wave rightward and compresses the PDR. A dense compressed region with a temperature about 30 K can be seen behind the shock front at $t$ = 3 Myr. The shock fronts are located at $z=$ 2.6 pc and $z=$ 5.8 pc at $t=$ 1 Myr and $t=$ 3 Myr, respectively. From these values, the propagation speed of the shock wave is 1.6 km s$^{-1}$. 
The reference model with larger chemical network shows different evolution from PDR benchmark model V2. 
The photoevaporation flow in the reference model has higher temperature of $\sim$ 1000 K and 
higher velocity of $\sim$ 9 km s$^{-1}$.
The propagation speed of the shock wave is 2.0 km s$^{-1}$. 
The stronger shock wave forms denser compressed region behind the shock front.  
 
\begin{figure}
   \begin{center}
   % color figure for online version
   \includegraphics[width=0.8 \columnwidth]{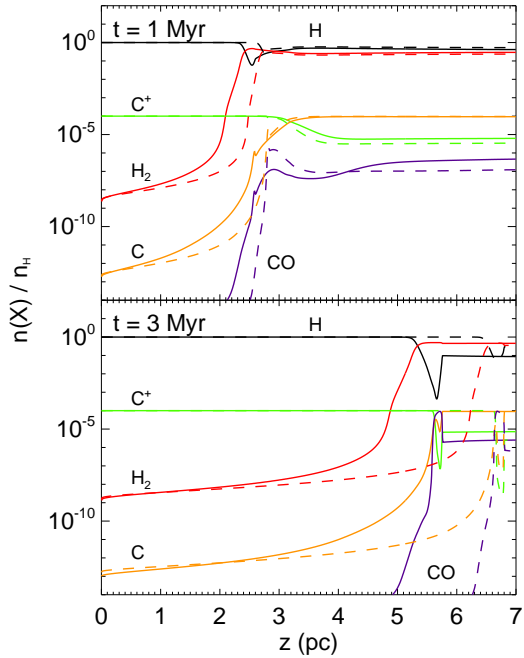}
   
   % black and white figure for print version
   %\includegraphics[width=0.8 \columnwidth]{f11.eps}
   \end{center}
   \caption{Chemical abundances at the time $t=1$ Myr (top) and $t=3$ Myr (bottom) as a function of z-coordinates in our hydrochemical simulation. The solid and the dashed lines represent results for PDR benchmark model V2 and for the reference model with larger chemical network, respectively.}
 \label{figure abundance_hydro}
 \end{figure}

Figure \ref{figure abundance_hydro} shows the fractional abundances of H$_2$, H, C$^+$, C, and CO at $t=$ 1 Myr and $t=$ 3 Myr. The molecular and atomic abundances clearly show the position of the shock front at both times, especially by the steep increase in abundance of CO. Since we assumed that the initial abundances were atomic, we can see the progressive increase of CO abundance in the inner shielded layers of the
cloud in time; and, similarly, the decrease of H due to conversion into H$_2$. Figure \ref{figure abundance_hydro} also clearly shows how the increased density at the shock front enhances the formation of CO, which for $t$ = 3 Myr is $\sim 2$ orders of magnitude higher than in the inner shielded layers. This result suggests that this contrast in abundance in CO may trace the position of the edge of the photoevaporation flow. The progression of the photo-evaporated flow is also traced, although not so dramatically, by the point of conversion of atomic into molecular gas, from H to H$_2$. Figure \ref{figure abundance_hydro} also shows that the shock wave traversed the entire computational domain before chemical equilibrium was reached in the uncompressed region. The time scale of hydrodynamic evolution is shorter than the time needed for chemical equilibrium. This affects the comparison with the models with variable temperature in Section~\ref{sect:pdr_benchmark}.
  
\begin{figure}
   \begin{center}
   % color figure for online version
   \includegraphics[width=0.8 \columnwidth]{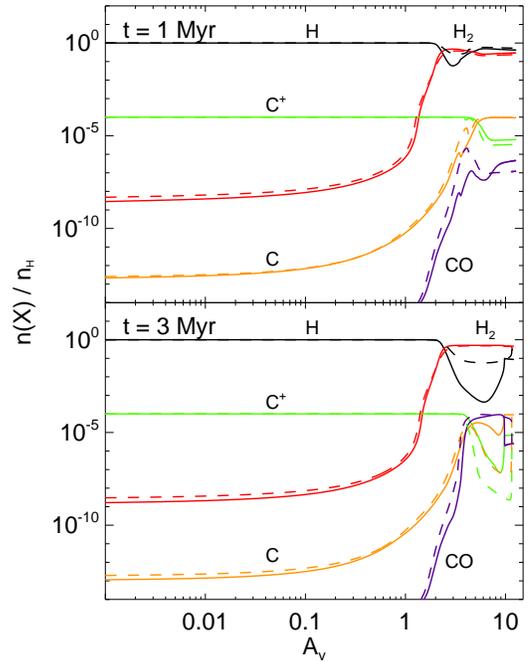}

   % black and white figure for print version
   %\includegraphics[width=0.8 \columnwidth]{f12.eps}
   \end{center}
   \caption{Same as Figure \ref{figure abundance_hydro}, except that the data are plotted as a function 
   of visual extinction.}
 \label{figure abundance_hydro_av}
 \end{figure}

For comparison with results of model V2 shown in Figure \ref{Vmodels}, we plot the fractional abundances of H$_2$, H, C$^+$, C, and CO at $t=$ 1 Myr and $t=$ 3 Myr as a function of visual extinction in Figure \ref{figure abundance_hydro_av}. 
The original V2 models were equivalent to the time when chemical equilibrium, in our case after 30 Myr. This difference in time accounts for the different abundances in the inner
shielded layers. We can see in Figure \ref{figure abundance_hydro_av} how the abundances at $A_V \la 0.1$ are markedly smaller, by an order of magnitude or more, for H$_2$, and C. After the passing of the photo-evaporating front, the density of this region is lower than that of the V2 case, which is reflected in lower shielding and a slower chemistry. On the other hand, at $A_V \ga 3$ the density is higher than model V2 by an order of magnitude due to shock compression and the effect of the higher gas density on CO, C, and C$^+$, both in terms of shielding and shorter chemical timescales can be seen clearly. The dissociation front of CO is located at $A_V \simeq 4$, while it is located at $A_V \simeq 10$ in model V2.
The reference model shows different chemical evolution from PDR benchmark model V2 in some respects. 
One difference is faster conversion from C into CO in the shock compressed region because of
higher density. 
Another difference is higher fractional abundance of H in the shock compressed region. 
Chemical network for the reference model contains more reactions producing atomic hydrogen, such as collisional dissociation or cosmic ray dissociation of molecules including hydrogen.
These reactions contributes higher fractional abundance of H. 

These hydrochemical hybrid simulations require about 40 times longer computation time than pure chemical simulations without the evolution of the hydrodynamics. This longer computation time is mainly due to its shorter time steps determined by Equation (\ref{dt hydro}). Computation times for chemical reactions, thermal processes, hydrodynamics, and radiative transfer account for 79.5, 18.5, 1.2, and 0.1 percent of the total computation time, respectively.

%%%%%%%%%%%%%%%%%%%%%%%%%%%%%%%%%%%%%%%%%%%%%%%%%%%%%%%%%%%%%%%%%%%%%%%%%%%%%%%%%%
%      discussion & conclusions
%%%%%%%%%%%%%%%%%%%%%%%%%%%%%%%%%%%%%%%%%%%%%%%%%%%%%%%%%%%%%%%%%%%%%%%%%%%%%%%%%%

\section{Discussion and Summary}\label{sec: conclusions}

The basic concepts and algorithms that had been adopted into the hydrochemical hybrid code, {\it KM2}, have been presented in this paper. The code is modular, and consists of a hydrodynamics module and a chemistry module. These modules have been verified with hydrodynamic test problems and PDR benchmark models that have been known in the respective fields for their performances. In particular, the method of time integration is advantageous in reducing the computation time that would otherwise be prohibitive for most problems. Because the time steps for chemical and thermal updates are locally determined at each computational cell, and large time steps can be used in cells where the gas is close to thermal equilibrium.
%Although a chemical network including only 287 reactions was used for benchmark models in Section \ref{sec: chem test}, the code can work with larger complex chemical networks. 
%Simulations with a full UMIST chemical network including 
%6173 reactions are consistent with those in Section \ref{sec: chem test}. 

As an application example of the {\it KM2} code, the first hydrochemical simulation of a photoevaporating PDR is presented in this paper. The photoevaporation flow changed the 
structure of PDR within a time scale shorter than the timescale needed to reach the chemical equilibrium, that is the timescale that is shorter would happen first before the chemical equilibria are reached locally. Profiles of chemical abundances were different from models without hydrodynamics. 
Moreover, photoevaporating PDRs with different chemical networks showed different thermal and 
hydrodynamical evolutions. These suggest that hydrochemical hybrid models are important for studies of photoevaporation where multiple timescales are involved and physical processes are interacting with one another. 
In addition to the PDRs, protoplanetary disks irradiated by central star or FUV fields may be important targets affected by photoevaporation in which various timescales are competitive.
Hydrochemical evolution models of photoevaporating PDRs and protoplanetary disks will be investigated in separate papers.

To apply the {\it KM2} code to more astrophysical problems, some extensions of its capabilities are planned as future work.
In this paper, the chemical networks include only gas-phase reactions, while grain-surface reactions also play important role in chemical evolution in dense molecular clouds. 
Inclusion of the grain-surface reactions will be desirable without the modification of the basic framework.
The code can be parallelized by message passing interface (MPI) library for large scale simulations. Reducing the computation time in the computation of the chemical reactions is also essential for large scale simulations.
As computations for chemical updates are done with local physical quantities, namely the density, temperature, and FUV intensity, node communication is not required in MPI parallelization of the chemistry solver. Good scalability and improvement of performance can be expected once the MPI is implemented.

Finally, this paper has demonstrated the essential implementation method for building a hybrid hydro-chemical code. This critical implementation allows the hydrodynamical and chemical evolution to be coupled by a straightforward operator splitting method in the {\it KM2} code. The implementation method can be easily extended and applied to other hydrodynamical codes for their complete hybridization of the thermal evolution at each time step.

\acknowledgments
The work has been funded and supported by the "CHARMS" Initiative under the Theoretical Institute for Advanced Research in Astrophysics (TIARA) of Academia Sinica Institute of Astronomy and Astrophysics (ASIAA). Numerical simulations were carried out in part on PC clusters at the Center for Computational Astrophysics, National Astronomical Observatory of Japan (NAOJ).

\bibliographystyle{apj} % style apj.bst
\bibliography{references}

\end{document}